\documentclass[twocolumn,preprintnumbers,showkeys,amsmath,amssymb]{revtex4}
\usepackage{graphicx}
\usepackage{amssymb}

\begin{document}
\title{Unconditionally secure quantum coin flipping}
\author{Guang Ping He}
\email{hegp@mail.sysu.edu.cn}
\affiliation{School of Physics, Sun Yat-sen University, Guangzhou
510275, China}

\begin{abstract}
Quantum coin flipping (QCF) is an essential primitive for quantum
cryptography. Unconditionally secure strong QCF with an arbitrarily small
bias was widely believed to be impossible. But basing on a problem which
cannot be solved without quantum algorithm, here we propose such a QCF
protocol, and show how it manages to evade all existing no-go proofs on QCF.
\end{abstract}


\maketitle



\section{Introduction}

Quantum coin flipping (QCF) \cite{qi365}, a.k.a. quantum coin tossing, is
aimed to provide a method for two separated parties Alice and Bob to
generate a random bit value $c=0$ or $1$ remotely, while they do not trust
each other. If the parties have opposite desired values, e.g., Alice wants $%
c=0$ while Bob wants $c=1$, then it is called weak QCF. Or if their desired
values are random, then it is called strong QCF. Here we concentrate on
strong QCF only. Such a QCF protocol is considered secure if neither party
can bias the outcome, so that $c=0$ and $c=1$ will both occur with the equal
probability $1/2$, just as if they are tossing an ideal fair coin.

QCF is an essential element of cheat-sensitive protocols \cite%
{qi150,qi442,HeSR}. It is also closely related with quantum bit commitment
(QBC) and quantum oblivious transfer (QOT) \cite{qbc121}, which are the
building blocks for more complicated quantum multi-party secure computation
protocols \cite{qi139}. However, it is widely believed that unconditionally
secure QCF is impossible. More rigorously, let $\varepsilon $ denote the
bias of a QCF protocol, such that the dishonest party can force the outcome $%
c=0$ or $c=1$ to occur with probability $1/2+\varepsilon >1/2$.
Unconditional security requires that $\varepsilon $ should be able to be
made arbitrarily small by increasing some parameters in the protocol,
without relying on any computational assumption or experimental constraint.
But there were proofs claiming that this can never be achieved. Instead, a
lower bound of the bias exists, which is $\varepsilon \geq 1/\sqrt{2}-1/2$
\cite{qbc121}, \cite%
{qi58,qbc46,qi145,qi146,qi76,qi817,qbc37,qbc54,qi246,qbc132,qbc19,qbc106,qbc112}%
. These no-go proofs are considered as casting very serious doubt on the
security of quantum cryptography in the so-called \textquotedblleft
post-cold-war\textquotedblright\ applications \cite{qi58}. All previously
proposed QCF protocols (e.g., Refs. \cite%
{qi132,qi148,qbc29,qbc76,qbc81,qbc117,qbc131}\ and the references therein)
are, unfortunately, limited by this bound.

Nevertheless, we find that this negative result is not sufficiently general
to cover all QCF protocols. Here we will propose an unconditionally secure
QCF protocol, and show one-by-one why the no-go proofs \cite{qbc121}, \cite%
{qi58,qbc46,qi145,qi146,qi76,qi817,qbc37,qbc54,qi246,qbc132,qbc19,qbc106,qbc112}
fail to apply to our protocol.

\section{The lie-detecting problem}

Since it is an important theoretical problem whether unconditionally secure
QCF exists in principle, here for simplicity, we only consider the ideal
case without practical imperfections, such as transmission errors, detection
loss or dark counts, etc.

Let us denote the state of a qubit as $\left\vert p,q\right\rangle $, where $%
p=0,1$ indicates the basis while $q=0,1$ distinguishes the two orthogonal
states in the same basis. That is, $\left\vert 0,0\right\rangle $ and $%
\left\vert 0,1\right\rangle $\ are the eigenstates of the $p=0$\ basis,
while $\left\vert 1,0\right\rangle \equiv (\left\vert 0,0\right\rangle
+\left\vert 0,1\right\rangle )/\sqrt{2}$ and $\left\vert 1,1\right\rangle
\equiv (\left\vert 0,0\right\rangle -\left\vert 0,1\right\rangle )/\sqrt{2}$%
\ are the eigenstates of the $p=1$\ basis. Our protocol is built around the
solutions to the following problem.

\bigskip

\textit{Lie-detecting Problem}\textsl{:}

Alice sends Bob $s$ qubits $\beta _{i}$ ($i\in S\equiv \{1,...,s\}$). Then
Bob announces his \textquotedblleft fake\textquotedblright\ measurement
result $\left\vert p_{i}^{\prime \prime },q_{i}^{\prime \prime
}\right\rangle _{\beta }\in \{\left\vert 0,0\right\rangle ,\left\vert
1,0\right\rangle ,\left\vert 0,1\right\rangle ,\left\vert 1,1\right\rangle
\} $ for each $\beta _{i}$, which is allowed to be different from his actual
measurement result $\left\vert p_{i}^{\prime },q_{i}^{\prime }\right\rangle
_{\beta }$. That is, it can be either of the following three types of lies:
\begin{eqnarray}
type\ a\ lies:\ p_{i}^{\prime \prime }=p_{i}^{\prime }\wedge q_{i}^{\prime
\prime }=\urcorner q_{i}^{\prime };  \nonumber \\
type\ b\ lies:\ p_{i}^{\prime \prime }=\urcorner p_{i}^{\prime }\wedge
q_{i}^{\prime \prime }=q_{i}^{\prime };  \nonumber \\
type\ c\ lies:\ p_{i}^{\prime \prime }=\urcorner p_{i}^{\prime }\wedge
q_{i}^{\prime \prime }=\urcorner q_{i}^{\prime }.
\end{eqnarray}


Now the question is: how many lies can Alice detect? More precisely, suppose
that types $a$, $b$ and $c$ lies occur with frequencies $f_{a}$, $f_{b}$ and
$f_{c}$, respectively. Our task is to express the total number of Alice's
detected lies with the parameters $f_{a}$, $f_{b}$, $f_{c}$ and $s$.

\bigskip

Note that it does not matter Bob measures each qubit before or after he
announces $\left\vert p_{i}^{\prime \prime },q_{i}^{\prime \prime
}\right\rangle _{\beta }$. Even if he announced without measuring it, he can
still measure it later and obtain $\left\vert p_{i}^{\prime },q_{i}^{\prime
}\right\rangle _{\beta }$, then compares it with $\left\vert p_{i}^{\prime
\prime },q_{i}^{\prime \prime }\right\rangle _{\beta }$ to learn the type of
lies that it belongs to. In this case, if $p_{i}^{\prime }$\ is randomly
chosen, then the values of $p_{i}^{\prime \prime }$ and $q_{i}^{\prime
\prime }$\ do not have a fixed relationship with $p_{i}^{\prime }$ and $%
q_{i}^{\prime }$, which is equivalent to announcing a type $a$, $b$ or $c$
lie or a honestly result with the equal probability $1/4$. Else if Bob
chooses $p_{i}^{\prime }=p_{i}^{\prime \prime }$\ ($p_{i}^{\prime }\neq
p_{i}^{\prime \prime }$), then the type $a$ lie and the honest result (the
type $b$ and $c$\ lies) occur with the equal probability $1/2$. Either way,
Bob is still able to control the values of $f_{a}$, $f_{b}$ and $f_{c}$.

Meanwhile, from Alice's point of view, some lies can be identified even
though Bob has not measured. For example, if Bob announces $\left\vert
p_{i}^{\prime \prime },q_{i}^{\prime \prime }\right\rangle _{\beta
}=\left\vert 0,0\right\rangle _{\beta }$ while Alice knows that she actually
sent $\left\vert 0,1\right\rangle _{\beta }$, she knows for sure that this $%
\left\vert p_{i}^{\prime \prime },q_{i}^{\prime \prime }\right\rangle
_{\beta }$ must be a lie no matter Bob has performed the measurement or not.
When Bob has not measured the corresponding $\beta _{i}$,
even if he later decides to measure it in the basis $p_{i}^{\prime
\prime }=0$\ honestly, he can only obtain $\left\vert p_{i}^{\prime
},q_{i}^{\prime }\right\rangle _{\beta }=\left\vert 0,1\right\rangle _{\beta
}$ so that his announced $\left\vert p_{i}^{\prime \prime },q_{i}^{\prime
\prime }\right\rangle _{\beta }$ ends up as a type $a$ lie. Else if he uses $%
p_{i}^{\prime \prime }=1$ as the measurement basis, then $\left\vert
p_{i}^{\prime \prime },q_{i}^{\prime \prime }\right\rangle _{\beta }$ will
inevitably become a type $b$ or $c$\ lie. In any case, he can never obtain
the result $\left\vert p_{i}^{\prime },q_{i}^{\prime }\right\rangle _{\beta
}=\left\vert 0,0\right\rangle _{\beta }$ so that he cannot make $\left\vert
p_{i}^{\prime \prime },q_{i}^{\prime \prime }\right\rangle _{\beta }$ an
honest announcement. Therefore, Alice can detect lies as usual even if Bob
delays his measurement.

\subsection{Algorithm I: the \textquotedblleft
semi-classical\textquotedblright\ algorithm}

A very intuitive and ordinary solution of the problem is as follows.

\textit{The state}: Alice determines the values of $p_{i}$ and $q_{i}$ ($%
i\in S$) beforehand, and prepares the initial state of each $\beta _{i}$ as
a pure state $\left\vert p_{i},q_{i}\right\rangle _{\beta }$ non-entangled
with any other system.

\textit{Lie-detecting strategy}: After Bob announced $\left\vert
p_{i}^{\prime \prime },q_{i}^{\prime \prime }\right\rangle _{\beta }$, Alice
compares it with $\left\vert p_{i},q_{i}\right\rangle _{\beta }$. Whenever
she finds $p_{i}^{\prime \prime }=p_{i}$ while $q_{i}^{\prime \prime
}=\urcorner q_{i}$, she knows that Bob told a lie. This is because in the
ideal setting, once Bob chose the basis $p_{i}^{\prime }$ correctly (i.e.,
it matches Alice's $p_{i}$), he should never find a wrong $q_{i}^{\prime }$
value in his measurement. Thus if he announces $\left\vert p_{i}^{\prime
},q_{i}^{\prime }\right\rangle _{\beta }$\ as $\left\vert p_{i}^{\prime
\prime },q_{i}^{\prime \prime }\right\rangle _{\beta }$ honestly, there
should not be $q_{i}^{\prime \prime }=\urcorner q_{i}$ while $p_{i}^{\prime
\prime }=p_{i}$. On the other hand, if $p_{i}^{\prime \prime }=\urcorner
p_{i}$ Alice will not be able to judge Bob's announcement. This is because
when Bob chose the wrong basis to measure the qubit, any result is possible
due to quantum uncertainty.

\textit{Result}: The total number of lies that Alice detected with this
algorithm can be evaluated as follows. Whenever Bob did not lie, Alice has
no way to claim that he did. So we need to concentrate only on the qubits
for which Bob indeed lies. Since Bob chooses the actual measurement basis $%
p_{i}^{\prime }$ randomly, for about $s/2$ qubits Bob will choose the
correct basis $p_{i}^{\prime }=p_{i}$ by chance. Among these qubits, all
type $a$ lies will inevitably be detected by Alice, while no types $b$ and $%
c $ can be judged for the reason stated above. Therefore Alice finds about $%
f_{a}s/2$\ lies. Meanwhile for the rest $s/2$ qubits which Bob measured with
the wrong basis $p_{i}^{\prime }=\urcorner p_{i}$, the probabilities of
finding $q_{i}^{\prime }=q_{i}$\ or $q_{i}^{\prime }=\urcorner q_{i}$\ are
both $1/2$. Therefore when Bob tells a type $b$ or $c$ lie, the announced
basis $p_{i}^{\prime \prime }=\urcorner p_{i}^{\prime }$ becomes the correct
basis so that the condition $p_{i}^{\prime \prime }=p_{i}$ is satisfied, and
$q_{i}^{\prime \prime }=\urcorner q_{i}$\ occurs with probability $1/2$.
Thus Alice finds about $(f_{b}+f_{c})(1/2)(s/2)$\ lies. But no $lie$ $a$
will be detected in this case since announcing $p_{i}^{\prime \prime
}=p_{i}^{\prime }$ will then be recognized by Alice as the wrong basis. All
in all, the total number of lies detected by Alice is about
\begin{equation}
l\backsim (\frac{1}{2}f_{a}+\frac{1}{4}f_{b}+\frac{1}{4}f_{c})s.  \label{eql}
\end{equation}%
Note that Alice needs not to know the values of $f_{a}$, $f_{b}$ and $f_{c}$%
. She simply follows the above lie-detecting strategy and this number can
automatically be reached.

This algorithm is called \textquotedblleft semi-classical\textquotedblright\
because the quantum property of the qubits is not fully utilized, except for
the quantum uncertainty involved in the measurement. Now we shall see that
if Alice further makes use of quantum entanglement, the problem can be
solved more efficiently.

Suppose that at the beginning, Alice introduces an ancillary system $\alpha
_{i}$ entangled with each $\beta _{i}$. She sends Bob $\beta _{i}$ and keeps
$\alpha _{i}$ at her side. Since Bob's actual result $\left\vert
p_{i}^{\prime },q_{i}^{\prime }\right\rangle _{\beta }$ has four possible
values $\left\vert 0,0\right\rangle $, $\left\vert 0,1\right\rangle $, $%
\left\vert 1,0\right\rangle $ and $\left\vert 1,1\right\rangle $, the
general form of the state of $\alpha _{i}\otimes \beta _{i}$ can be taken as%
\begin{eqnarray}
\left\vert \alpha _{i}\otimes \beta _{i}\right\rangle &=&c_{x}\left\vert
x\right\rangle _{\alpha }\otimes \left\vert 0,0\right\rangle _{\beta
}+c_{y}\left\vert y\right\rangle _{\alpha }\otimes \left\vert
1,0\right\rangle _{\beta }  \nonumber  \label{general} \\
&&+c_{x^{\prime }}\left\vert x^{\prime }\right\rangle _{\alpha }\otimes
\left\vert 0,1\right\rangle _{\beta }+c_{y^{\prime }}\left\vert y^{\prime
}\right\rangle _{\alpha }\otimes \left\vert 1,1\right\rangle _{\beta }.
\nonumber \\
&&
\end{eqnarray}%
By choosing different combinations of the superposition coefficients $c_{x}$%
, $c_{y}$, $c_{x^{\prime }}$ and $c_{y^{\prime }}$, we find that the
following three algorithms are especially useful for our current purpose.

\subsection{Algorithm II: quantum algorithm with maximally entangled states}

\textit{The state}: Alice takes
\begin{equation}
\left\vert \alpha _{i}\otimes \beta _{i}\right\rangle =(\left\vert
x\right\rangle _{\alpha }\otimes \left\vert 0,0\right\rangle _{\beta
}+\left\vert y\right\rangle _{\alpha }\otimes \left\vert 0,1\right\rangle
_{\beta })/\sqrt{2},  \label{eq1}
\end{equation}%
which is the maximally entangled state of $\alpha _{i}\otimes \beta _{i}$.
Here and in the following context $\left\vert x\right\rangle _{\alpha }$ and
$\left\vert y\right\rangle _{\alpha }$ are taken to be orthogonal to each
other.

\textit{Lie-detecting strategy}: After Bob announced $\left\vert
p_{i}^{\prime \prime },q_{i}^{\prime \prime }\right\rangle _{\beta }$, Alice
measures $\alpha _{i}$ in the basis which forces $\beta _{i}$ to collapse to
$p_{i}^{\prime \prime }$. (Note that the measurement of Alice and Bob on the
entangled state is permutable. Therefore, though the actual case could be
that Bob has measured $\beta _{i}$ and caused $\alpha _{i}$ to collapse
before Alice performs her measurement, it is equivalent to the case where
Alice has measured $\alpha _{i}$ and caused $\beta _{i}$ to collapse first.)
That is, if $p_{i}^{\prime \prime }=0$, she measures $\alpha _{i}$ in the
basis $\{\left\vert x\right\rangle _{\alpha },\left\vert y\right\rangle
_{\alpha }\}$ so that $\beta _{i}$ collapses to $\left\vert 0,0\right\rangle
_{\beta }$ or $\left\vert 0,1\right\rangle _{\beta }$. Else if $%
p_{i}^{\prime \prime }=1$, she measures $\alpha _{i}$ in the basis $%
\{(\left\vert x\right\rangle _{\alpha }+\left\vert y\right\rangle _{\alpha
})/\sqrt{2},(\left\vert x\right\rangle _{\alpha }-\left\vert y\right\rangle
_{\alpha })/\sqrt{2}\}$ so that $\beta _{i}$ collapses to $\left\vert
1,0\right\rangle _{\beta }$ or $\left\vert 1,1\right\rangle _{\beta }$. As a
result, the real \textquotedblleft initial\textquotedblright\ basis $p_{i}$
of the qubit $\beta _{i}$ always equals to Bob's announced basis $%
p_{i}^{\prime \prime }$. Then Alice knows that Bob told a lie whenever she
finds $q_{i}^{\prime \prime }=\urcorner q_{i}$.

\textit{Result}: Since $p_{i}$ always equals to $p_{i}^{\prime \prime }$ in
this algorithm, it can be seen that the probability for any type of lies to
be detected by Alice is doubled when comparing with that of the
\textquotedblleft semi-classical\textquotedblright\ algorithm, where $%
p_{i}=p_{i}^{\prime \prime }$\ occurs at half of the cases only. Therefore,
if Alice measures all $\alpha _{i}$'s in the $s$ pairs of entangled states,
the total number of lies that she can detect is $2l\thicksim
(f_{a}+f_{b}/2+f_{c}/2)s$. On the other hand, if she is required to detect $%
l $ (defined by Eq. (\ref{eql})) lies only, just as what can be achieved in
the \textquotedblleft semi-classical\textquotedblright\ algorithm, the
number of $\alpha _{i}$\ which she needs to measure is $s/2$ only. The other
$s/2$ pairs of entangled states can be left intact.

In fact, the \textquotedblleft semi-classical\textquotedblright\ algorithm
can be written in an equivalent form, in which the initial state is also
prepared as Eq. (\ref{eq1}), but Alice always picks a random basis $p_{i}$
for measuring $\alpha _{i}$, instead of choosing $p_{i}$ according to Bob's $%
p_{i}^{\prime \prime }$. Then to detect $l$ lies, she has to measure all the
$s$ entangled states. In this sense, the current quantum algorithm is more
efficient than the \textquotedblleft semi-classical\textquotedblright\ one
as less $\alpha _{i}$ is measured.

\subsection{Algorithm III: quantum algorithm with non-maximally entangled
states}

Although the above algorithm is the best we found for Alice to detect as
much lies as possible, it is not the most efficient one if only $l$ lies are
required to be detected. With other forms of $\left\vert \alpha _{i}\otimes
\beta _{i}\right\rangle $, the number of $\alpha _{i}$ that Alice needs to
measure can be further reduced.

\textit{The state}: Alice prepares
\begin{equation}
\left\vert \alpha _{i}\otimes \beta _{i}\right\rangle =\cos \theta
_{i}\left\vert x\right\rangle _{\alpha }\otimes \left\vert
0,q_{i}\right\rangle _{\beta }+\sin \theta _{i}\left\vert y\right\rangle
_{\alpha }\otimes \left\vert 1,q_{i}\right\rangle _{\beta },  \label{eq2}
\end{equation}%
where $q_{i}\in \{0,1\}$ and $0<\theta _{i}<\pi /2$. Note that $\theta _{i}$
needs not to be different for each $i$. For example, Alice can always take $%
\theta _{i}=\pi /4$. But to make our analysis sufficiently general to cover
all algorithms with the same property, in this subsection we do not limit $%
\theta _{i}$ to a fixed value.

\textit{Lie-detecting strategy}: After Bob announced $\left\vert
p_{i}^{\prime \prime },q_{i}^{\prime \prime }\right\rangle _{\beta }$, Alice
divides the $s$ pairs of entangled states\ into two subsets: Set $M$
includes all those satisfying $q_{i}^{\prime \prime }\neq q_{i}$, and set $U$
includes all those satisfying $q_{i}^{\prime \prime }=q_{i}$. When Alice's
goal is to detect $l$ lies only, she can simply leave all the qubits in $U$
unmeasured. Meanwhile, for these in $M$ she measures each $\alpha _{i}$ in
the basis $\{\left\vert x\right\rangle _{\alpha },\left\vert y\right\rangle
_{\alpha }\}$. She sets $p_{i}=0$ ( $p_{i}=1$) if the measurement result is $%
\left\vert x\right\rangle _{\alpha }$ ($\left\vert y\right\rangle _{\alpha }$%
). Then she sets $L=\{i\in M|p_{i}=p_{i}^{\prime \prime }\}$ as the set of
qubits for which she detected lies.

\textit{Result}: After applying this strategy to all $\alpha _{i}\otimes
\beta _{i}$ ($i\in S$), set $S$ is divided into three subsets:

$U\equiv \{i\in S|q_{i}^{\prime \prime }=q_{i}\}$, for which Alice has not
measured $\alpha _{i}$,

$L\equiv \{i\in S|(q_{i}^{\prime \prime }\neq q_{i})\wedge (p_{i}^{\prime
\prime }=p_{i})\}$, for which Alice has measured $\alpha _{i}$ and detected
that $\left\vert p_{i}^{\prime \prime },q_{i}^{\prime \prime }\right\rangle
_{\beta }$\ is a lie, and

$N\equiv \{i\in S|(q_{i}^{\prime \prime }\neq q_{i})\wedge (p_{i}^{\prime
\prime }\neq p_{i})\}$, for which Alice has measured $\alpha _{i}$ and she
cannot judge whether $\left\vert p_{i}^{\prime \prime },q_{i}^{\prime \prime
}\right\rangle _{\beta }$ is honest or a lie.

In addition, $M=L\cup N=\{i\in S|q_{i}^{\prime \prime }\neq q_{i}\}$.

Here we will show that set $N$ has a distinguishing property, that it does
not contain any type $b$ lie. Suppose that Alice takes $q_{i}=0$. In this
case Eq. (\ref{eq2}) becomes
\begin{equation}
\left\vert \alpha _{i}\otimes \beta _{i}\right\rangle =\cos \theta
_{i}\left\vert x\right\rangle _{\alpha }\otimes \left\vert 0,0\right\rangle
_{\beta }+\sin \theta _{i}\left\vert y\right\rangle _{\alpha }\otimes
\left\vert 1,0\right\rangle _{\beta },
\end{equation}%
which can also be written as
\begin{eqnarray}
\left\vert \alpha _{i}\otimes \beta _{i}\right\rangle &=&(\cos \theta
_{i}\left\vert x\right\rangle _{\alpha }+\frac{\sin \theta _{i}}{\sqrt{2}}%
\left\vert y\right\rangle _{\alpha })\otimes \left\vert 0,0\right\rangle
_{\beta }  \nonumber \\
&&+\frac{\sin \theta _{i}}{\sqrt{2}}\left\vert y\right\rangle _{\alpha
}\otimes \left\vert 0,1\right\rangle _{\beta },  \label{basis0}
\end{eqnarray}%
or
\begin{eqnarray}
\left\vert \alpha _{i}\otimes \beta _{i}\right\rangle &=&(\frac{\cos \theta
_{i}}{\sqrt{2}}\left\vert x\right\rangle _{\alpha }+\sin \theta
_{i}\left\vert y\right\rangle _{\alpha })\otimes \left\vert 1,0\right\rangle
_{\beta }  \nonumber \\
&&+\frac{\cos \theta _{i}}{\sqrt{2}}\left\vert x\right\rangle _{\alpha
}\otimes \left\vert 1,1\right\rangle _{\beta }.  \label{basis1}
\end{eqnarray}%
According to the lie-detecting strategy, Alice will measure $\alpha _{i}$ in
the basis $\{\left\vert x\right\rangle _{\alpha },\left\vert y\right\rangle
_{\alpha }\}$ only if Bob announces $q_{i}^{\prime \prime }=1$. In this
case, if $p_{i}^{\prime \prime }=0$, i.e., Bob announces $\left\vert
p_{i}^{\prime \prime },q_{i}^{\prime \prime }\right\rangle _{\beta
}=\left\vert 0,1\right\rangle _{\beta }$, Eq. (\ref{basis0}) shows that when
this is the honest result, i.e., Bob indeed measured $\beta _{i}$ in the $%
p_{i}^{\prime }=0$\ basis and finds $\left\vert p_{i}^{\prime
},q_{i}^{\prime }\right\rangle _{\beta }=\left\vert 0,1\right\rangle _{\beta
}$, then $\alpha _{i}$ should collapse to $\left\vert y\right\rangle
_{\alpha }$. Therefore, if Alice's measurement result on $\alpha _{i}$ is $%
\left\vert x\right\rangle _{\alpha }$, she knows for sure that $\left\vert
p_{i}^{\prime \prime },q_{i}^{\prime \prime }\right\rangle _{\beta
}=\left\vert 0,1\right\rangle _{\beta }$ is a lie. Especially, if this $%
\left\vert p_{i}^{\prime \prime },q_{i}^{\prime \prime }\right\rangle
_{\beta }$ is a type $b$\ lie, i.e., Bob actually measured $\beta _{i}$ in
the $p_{i}^{\prime }=1$\ basis and the actual result is $\left\vert
p_{i}^{\prime },q_{i}^{\prime }\right\rangle _{\beta }=\left\vert
1,1\right\rangle _{\beta }$, Eq. (\ref{basis1}) shows that $\alpha _{i}$
will collapse to $\left\vert x\right\rangle _{\alpha }$ with probability $1$
so that Alice can always detect it and include it in set $L$. Consequently,
if Alice measures $\alpha _{i}$ and finds that the result is $\left\vert
y\right\rangle _{\alpha }$, she knows with certainty that $\left\vert
p_{i}^{\prime \prime },q_{i}^{\prime \prime }\right\rangle _{\beta
}=\left\vert 0,1\right\rangle _{\beta }$ is not a type $b$ lie.

On the other hand, if Bob announces $\left\vert p_{i}^{\prime \prime
},q_{i}^{\prime \prime }\right\rangle _{\beta }=\left\vert 1,1\right\rangle
_{\beta }$, Eq. (\ref{basis1}) shows that when $\left\vert p_{i}^{\prime
\prime },q_{i}^{\prime \prime }\right\rangle _{\beta }$ is the honest
result, then $\alpha _{i}$ should collapse to $\left\vert x\right\rangle
_{\alpha }$. Therefore, if Alice measures $\alpha _{i}$ and finds that the
result is $\left\vert y\right\rangle _{\alpha }$, she knows for sure that $%
\left\vert p_{i}^{\prime \prime },q_{i}^{\prime \prime }\right\rangle
_{\beta }=\left\vert 1,1\right\rangle _{\beta }$ is a lie. Again, if this $%
\left\vert p_{i}^{\prime \prime },q_{i}^{\prime \prime }\right\rangle
_{\beta }$ is a type $b$\ lie, i.e., Bob actually measured $\beta _{i}$ in
the $p_{i}^{\prime }=0$\ basis and the actual result is $\left\vert
p_{i}^{\prime },q_{i}^{\prime }\right\rangle _{\beta }=\left\vert
0,1\right\rangle _{\beta }$, Eq. (\ref{basis0}) shows that $\alpha _{i}$
will collapse to $\left\vert y\right\rangle _{\alpha }$ with probability $1$
so that Alice can always detect it and include it in set $L$. Consequently,
if Alice measures $\alpha _{i}$ and finds that the result is $\left\vert
x\right\rangle _{\alpha }$, she knows with certainty that $\left\vert
p_{i}^{\prime \prime },q_{i}^{\prime \prime }\right\rangle _{\beta
}=\left\vert 1,1\right\rangle _{\beta }$ is not a type $b$ lie.

The case where Alice takes $q_{i}=1$ can also be analyzed similarly. Once
again, if Bob announces $\left\vert p_{i}^{\prime \prime },q_{i}^{\prime
\prime }\right\rangle _{\beta }=\left\vert 0,0\right\rangle _{\beta }$ ($%
\left\vert p_{i}^{\prime \prime },q_{i}^{\prime \prime }\right\rangle
_{\beta }=\left\vert 1,0\right\rangle _{\beta }$) and Alice's measurement
result on $\alpha _{i}$ is $\left\vert y\right\rangle _{\alpha }$ ($%
\left\vert x\right\rangle _{\alpha }$), she knows with certainty that $%
\left\vert p_{i}^{\prime \prime },q_{i}^{\prime \prime }\right\rangle
_{\beta }$ is not a type $b$ lie. Otherwise, she knows that it must be a
lie. Namely, all type $b$ lies in set $M$ will be detected and put into set $%
L$, so that set $N$ will contain no type $b$ lie at all.

The sizes of sets $U$, $L$, $M$ and $N$ can then be evaluated as follows.
Denote the frequency that Bob announces the honest results as%
\begin{equation}
f_{h}\equiv 1-f_{a}-f_{b}-f_{c}.  \label{fh}
\end{equation}%
Eq. (\ref{basis0}) shows that if Bob measures $\beta _{i}$\ in the $%
p_{i}^{\prime }=0$ basis, the result $q_{i}^{\prime }\neq q_{i}$\ will occur
with probability $(\sin ^{2}\theta _{i})/2$, while Eq. (\ref{basis1}) shows
that if Bob measures $\beta _{i}$\ in the $p_{i}^{\prime }=1$ basis, the
result $q_{i}^{\prime }\neq q_{i}$\ will occur with probability $(\cos
^{2}\theta _{i})/2$. Thus the average probability for him to find $%
q_{i}^{\prime }\neq q_{i}$\ is $((\sin ^{2}\theta _{i})/2+(\cos ^{2}\theta
_{i})/2)/2=1/4$. For types $a$ and $c$ lies, Bob always announces $%
q_{i}^{\prime \prime }\neq q_{i}^{\prime }$. Thus they will satisfy $%
q_{i}^{\prime \prime }=q_{i}$\ and fall into set $U$ with probability $1/4$.
For honest results and type $b$ lies, Bob always announces $q_{i}^{\prime
\prime }=q_{i}^{\prime }$. Thus they will satisfy $q_{i}^{\prime \prime
}=q_{i}$\ and fall into set $U$ with probability $3/4$. Therefore, the size
of $U$ is%
\begin{equation}
\left\vert U\right\vert =(\frac{3}{4}f_{h}+\frac{1}{4}f_{a}+\frac{3}{4}f_{b}+%
\frac{1}{4}f_{c})s.  \label{sizeU}
\end{equation}

Now consider set $L$. Obviously, if Bob announces the result honestly, Alice
will not detect it as a lie. Thus honest results will not present in $L$. We
also showed above that set $N$ will not contain any type $b$ lie. Therefore,
all the type $b$ lies not included in $U$ will present in $L$, so that the
number is $f_{b}s/4$. Now suppose that $\left\vert p_{i}^{\prime \prime
},q_{i}^{\prime \prime }\right\rangle _{\beta }=\left\vert 0,1\right\rangle
_{\beta }$ is a type $a$ lie and Alice has chosen $q_{i}=0$ to prepare the
state $\left\vert \alpha _{i}\otimes \beta _{i}\right\rangle $. That is,
Bob's actual result is $\left\vert p_{i}^{\prime },q_{i}^{\prime
}\right\rangle _{\beta }=\left\vert 0,0\right\rangle _{\beta }$. As we
showed above, if Alice finds $\left\vert x\right\rangle _{\alpha }$ in her
measurement on $\alpha _{i}$ then she detects $\left\vert p_{i}^{\prime
\prime },q_{i}^{\prime \prime }\right\rangle _{\beta }=\left\vert
0,1\right\rangle _{\beta }$ as a lie. Eq. (\ref{basis0}) shows that the case
where Alice finds $\left\vert x\right\rangle _{\alpha }$ while Bob finds $%
\left\vert 0,0\right\rangle _{\beta }$ occurs\ with probability $\cos
^{2}\theta _{i}$. On the other hand, when $\left\vert p_{i}^{\prime \prime
},q_{i}^{\prime \prime }\right\rangle _{\beta }=\left\vert 1,1\right\rangle
_{\beta }$ is a type $a$ lie and Alice has chosen $q_{i}=0$, Eq. (\ref%
{basis1}) shows that Alice will detect it (i.e., she finds $\left\vert
y\right\rangle _{\alpha }$ while Bob's actual result is $\left\vert
1,0\right\rangle _{\beta }$)\ with probability $\sin ^{2}\theta _{i}$. In
average, Bob's type $a$ lies will be detected with probability $(\cos
^{2}\theta _{i}+\sin ^{2}\theta _{i})/2=1/2$. Analyzing the $q_{i}=1$ case
will also give the similar result. Consequently, the number of type $a$ lies
in set $L$ is $f_{a}s/2$. Finally, when $q_{i}=0$, if $\left\vert
p_{i}^{\prime \prime },q_{i}^{\prime \prime }\right\rangle _{\beta
}=\left\vert 0,1\right\rangle _{\beta }$ is a type $c$ lie, Eq. (\ref{basis1}%
) shows that Alice will detect it (i.e., she finds $\left\vert
x\right\rangle _{\alpha }$ while Bob's actual result is $\left\vert
1,0\right\rangle _{\beta }$)\ with probability $(\cos ^{2}\theta _{i})/2$.
Else if $\left\vert p_{i}^{\prime \prime },q_{i}^{\prime \prime
}\right\rangle _{\beta }=\left\vert 1,1\right\rangle _{\beta }$ is a type $c$
lie, Eq. (\ref{basis0}) shows that Alice will detect it (i.e., she finds $%
\left\vert y\right\rangle _{\alpha }$ while Bob's actual result is $%
\left\vert 0,0\right\rangle _{\beta }$)\ with probability $(\sin ^{2}\theta
_{i})/2$. In average, Bob's type $c$ lies will be detected with probability $%
((\cos ^{2}\theta _{i})/2+(\sin ^{2}\theta _{i})/2)/2=1/4$. So does the $%
q_{i}=1$ case. That is, the number of type $c$ lies in set $L$ is $f_{c}s/4$%
. Thus we know that the size of $L$ is%
\begin{equation}
\left\vert L\right\vert =(\frac{1}{2}f_{a}+\frac{1}{4}f_{b}+\frac{1}{4}%
f_{c})s=l.  \label{sizeL}
\end{equation}%
As a consequence, the size of $N$ is%
\begin{equation}
\left\vert N\right\vert =s-\left\vert U\right\vert -\left\vert L\right\vert
=(\frac{1}{4}f_{h}+\frac{1}{4}f_{a}+\frac{1}{2}f_{c})s,  \label{sizeN}
\end{equation}%
and the size of $M$ is%
\begin{equation}
\left\vert M\right\vert =\left\vert L\right\vert +\left\vert N\right\vert =(%
\frac{1}{4}f_{h}+\frac{3}{4}f_{a}+\frac{1}{4}f_{b}+\frac{3}{4}f_{c})s.
\label{sizeM}
\end{equation}%
Combining with Eq. (\ref{fh}), we have%
\begin{equation}
\left\vert M\right\vert =(\frac{1}{4}+\frac{f_{a}+f_{c}}{2})s.
\label{sizeM'}
\end{equation}%
In Fig.1 we summarized the above sizes and properties of these sets.


\begin{figure*}[tbp]
\centering
\includegraphics{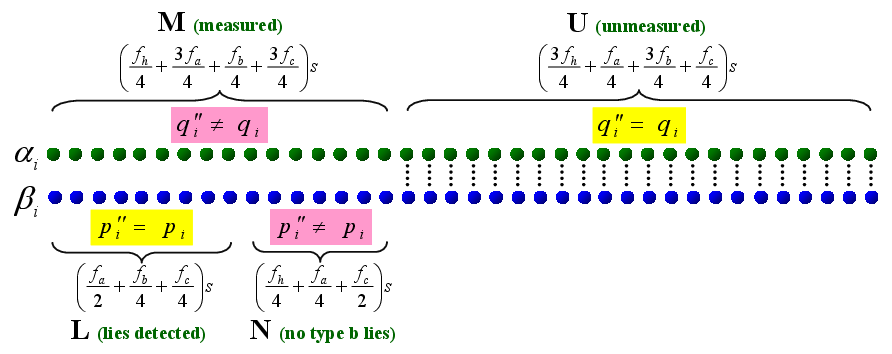}
\caption{Resultant subsets of the qubits after applying the lie-detecting
algorithm III. Note that for illustration purposes, the qubits are regrouped
in this diagram according to the subsets that they belong to, instead of
following their original order indicated by the index $i$.}
\label{fig:epsart}
\end{figure*}


Thus we see that Alice can indeed detect $l$ lies, i.e., achieve the same
goal of the \textquotedblleft semi-classical\textquotedblright\ algorithm,
while only about $\left\vert M\right\vert =[1/4+(f_{a}+f_{c})/2]s$\ qubits
are measured. This number is even less than that of algorithm II as long as $%
f_{a}+f_{c}<1/2$.

\subsection{Algorithm IV: alternative quantum algorithm with non-maximally
entangled states}

Similarly, we have the following algorithm.

\textit{The state}: Alice prepares
\begin{equation}
\left\vert \alpha _{i}\otimes \beta _{i}\right\rangle =\cos \theta
_{i}\left\vert x\right\rangle _{\alpha }\otimes \left\vert
0,q_{i}\right\rangle _{\beta }+\sin \theta _{i}\left\vert y\right\rangle
_{\alpha }\otimes \left\vert 1,\urcorner q_{i}\right\rangle _{\beta }.
\label{eq3}
\end{equation}

\textit{Lie-detecting strategy}: After Bob announced $\left\vert
p_{i}^{\prime \prime },q_{i}^{\prime \prime }\right\rangle _{\beta }$, Alice
divides the $s$ pairs of entangled states\ into two subsets: Set $M^{\prime
} $ includes all those satisfying either $\left\vert p_{i}^{\prime \prime
},q_{i}^{\prime \prime }\right\rangle _{\beta }=\left\vert 0,\urcorner
q_{i}\right\rangle _{\beta }$ or $\left\vert p_{i}^{\prime \prime
},q_{i}^{\prime \prime }\right\rangle _{\beta }=\left\vert
1,q_{i}\right\rangle _{\beta }$, and set $U^{\prime }\equiv S-M^{\prime }$.
Again, she only measures each $\alpha _{i}$ ($i\in M^{\prime }$) in the
basis $\{\left\vert x\right\rangle _{\alpha },\left\vert y\right\rangle
_{\alpha }\}$, and sets $p_{i}=0$ ( $p_{i}=1$) if she finds $\left\vert
x\right\rangle _{\alpha }$ ($\left\vert y\right\rangle _{\alpha }$). Then
she sets $L^{\prime }=\{i\in M^{\prime }\wedge p_{i}=p_{i}^{\prime \prime
}\} $ as the set of qubits for which she detected lies.

\textit{Result}: With a similar analysis to that of algorithm III, it can be
proven that set $N^{\prime }\equiv M^{\prime }-L^{\prime }$ does not contain
any type $c$\ lie, and Alice needs to measure about $\left\vert M^{\prime
}\right\vert =[1/4+(f_{a}+f_{b})/2]s$ qubits to detect $l$ lies.

This number is also less than that of algorithm II as long as $%
f_{a}+f_{b}<1/2$. But whether this algorithm is more efficient than
algorithm III or not will depend on the comparison between $f_{b}$\ and $%
f_{c}$, which is determined by Bob. In the following we will use the case $%
f_{b}>f_{c}$\ as an example and build a QCF protocol upon algorithm III. But
in fact algorithm IV can also be used to build a similar protocol in the
case $f_{b}<f_{c}$.

%
%
%
%
%
%
%

\section{Our protocol}

Basing on the above lie-detecting problem and algorithm III, we build the
following protocol.

\bigskip

\textit{The QCF protocol (for generating a random bit }$c$\textit{):}

(1) Alice and Bob agree on a subset $C$\ of $s$-bit strings, i.e., $C\subset
\{0,1\}^{s}$, which has the size $\left\vert C\right\vert \backsim 2^{k}$ ($%
k<s$). Also, the elements of $C$ (called codewords) should satisfy both of
the following requirements:

\qquad (i) The numbers of codewords having odd and even parity should both
be non-trivial.

\qquad (ii) The distance (i.e., the number of different bits) between any
two codewords is not less than $d$ ($d<s/4$).

A binary linear $(s,k,d)$-code (or its selected subset, as we do not need it
to meet all the requirements of linear classical error-correction code
except the above two) generally fits the job.

(2) Alice chooses a codeword $q\equiv q_{1}q_{2}...q_{s}\in C$, and prepares
$s$ pairs of quantum systems $\alpha _{i}\otimes \beta _{i}$ ($i\in S\equiv
\{1,...,s\}$), each of which is in the state
\begin{equation}
\left\vert \alpha _{i}\otimes \beta _{i}\right\rangle =\frac{1}{\sqrt{2}}%
(\left\vert x\right\rangle _{\alpha }\otimes \left\vert 0,q_{i}\right\rangle
_{\beta }+\left\vert y\right\rangle _{\alpha }\otimes \left\vert
1,q_{i}\right\rangle _{\beta }),  \label{eq2'}
\end{equation}%
which is Eq. (\ref{eq2}) with $\theta _{i}$ fixed as $\pi /4$. She keeps
each $\alpha _{i}$ at her side, while sends $\beta _{i}$ to Bob.

(3) Bob announces his \textquotedblleft fake\textquotedblright\ measurement
result $\left\vert p_{i}^{\prime \prime },q_{i}^{\prime \prime
}\right\rangle _{\beta }$ for each $\beta _{i}$. Here the numbers of $0$'s
and $1$'s in $q^{\prime \prime }\equiv q_{1}^{\prime \prime }q_{2}^{\prime
\prime }...q_{s}^{\prime \prime }$\ should both be non-trivial. Also, types $%
a$, $b$ and $c$ lies should occur with frequencies $f_{a}$, $f_{b}$ and $%
f_{c}$ that satisfy $2d/s<f_{a}+f_{c}<1/2$ and $f_{b}>f_{c}$.

(4) Alice checks that the numbers of $0$'s and $1$'s in $q^{\prime \prime }$
are both non-trivial. Then she applies the lie-detecting strategy in
algorithm III. Namely, for each $\alpha _{i}$, Alice leaves it unmeasured if
$q_{i}^{\prime \prime }=q_{i}$. Else if $q_{i}^{\prime \prime }=\urcorner
q_{i}$, she measures $\alpha _{i}$ in the basis $\{\left\vert x\right\rangle
_{\alpha },\left\vert y\right\rangle _{\alpha }\}$, and sets $p_{i}=0$ ( $%
p_{i}=1$) if the result is $\left\vert x\right\rangle _{\alpha }$ ($%
\left\vert y\right\rangle _{\alpha }$). Then she divides set $S$ into three
subsets:

$\qquad U\equiv \{i\in S|q_{i}^{\prime \prime }=q_{i}\}$,

$\qquad L\equiv \{i\in S|(q_{i}^{\prime \prime }\neq q_{i})\wedge
(p_{i}^{\prime \prime }=p_{i})\}$,

$\qquad N\equiv \{i\in S|(q_{i}^{\prime \prime }\neq q_{i})\wedge
(p_{i}^{\prime \prime }\neq p_{i})\}$.

She also checks that the size of set $M=L\cup N$ satisfies $\left\vert
M\right\vert >d+s/4$\ (which guarantees that $2d/s<f_{a}+f_{c}$\ as can be
seen from Eq. (\ref{sizeM'})).

(5) Alice announces set $L$ to Bob.

(6) Bob checks that the size of $L$ satisfies Eq. (\ref{sizeL})
approximately (where a small amount of deviation is allowed, since the
equation was merely a statistical estimation), and all $\left\vert
p_{i}^{\prime \prime },q_{i}^{\prime \prime }\right\rangle _{\beta }$'s ($%
i\in L$) are indeed lies (with no deviation allowed).

(7) Bob announces a random bit $f\in \{0,1\}$.

(8) Alice announces sets $N$ and $U$ to Bob, and sends him all $\alpha _{i}$%
's in set $U$.

(9) From the definitions of $U$, $L$, $N$ Bob can deduce $q$ from his own $%
q^{\prime \prime }$. Then he checks that:

\qquad (9.1) $q$ is exactly a codeword of $C$.

\qquad (9.2) The sizes of $N$ and $U$ satisfy Eqs. (\ref{sizeN}) and (\ref%
{sizeU}) approximately. (A small amount of deviation is allowed.)

\qquad (9.3) Set $N$ does not contain type $b$ lies. (No deviation allowed.)

\qquad (9.4) He measures each $\alpha _{i}$ in set $U$ using any basis he
prefers, and compares the result with his $\left\vert p_{i}^{\prime
},q_{i}^{\prime }\right\rangle _{\beta }$ 
to verify that the state of $\alpha _{i}\otimes \beta _{i}$ consists with
both Eq. (\ref{eq2'}) and the $q_{i}$\ value that he deduced from Alice's
announced $U$.

(10) If the above security checks are passed, they take the coin-flip result
as $c=0\oplus f$\ ($c=1\oplus f$) if $\sum\nolimits_{i\in N}q_{i}$ is an
even (odd) number.

\bigskip

\section{Security proof}

We can see that the main idea of the protocol is as follows. Bob requires
Alice to solve the lie-detecting problem with the highest efficiency (i.e.,
the number of unmeasured $\alpha _{i}$ should be maximized) as he will
checks the size of $U$ and measures both $\alpha _{i}$ and $\beta _{i}$ for
any $i\in U$. Thus Alice is forced to apply algorithm III when Bob chooses $%
f_{a}+f_{c}<1/2$ and $f_{b}>f_{c}$ since other algorithms are less efficient
in this case. As a consequence, the qubits are divided into subsets $M$ and $%
U$ by the comparison between $q\equiv q_{1}q_{2}...q_{s}$\ and $q^{\prime
\prime }\equiv q_{1}^{\prime \prime }q_{2}^{\prime \prime }...q_{s}^{\prime
\prime }$, as shown in Fig.1. The coin-flip result $c$\ is essentially
determined by the $q_{i}$\ values of the qubits felt into the subset $N$.
Bob also generates a bit $f$ in step (7) which serves as a flip of the final
$c$, so that Alice's biasing the states before this step will be in vain as
different values of $f$ could bias the final towards opposite directions.
Therefore, to prove that the protocol is unconditionally secure, we only
need to prove that Bob does not know Alice's $q$ (and therefore $N$) before
step (7), while Alice cannot change $q$ (and therefore $N$) after step (7).

Consider Bob's case first. Eq. (\ref{eq2'}) indicates that Bob's own
operation that maximize his information on $q_{i}$ is the one that
distinguishes the reduced density matrix $\rho _{q_{i}=0}\equiv (\left\vert
0,0\right\rangle _{\beta }\left\langle 0,0\right\vert +\left\vert
1,0\right\rangle _{\beta }\left\langle 1,0\right\vert )/2$ from $\rho
_{q_{i}=1}\equiv (\left\vert 0,1\right\rangle _{\beta }\left\langle
0,1\right\vert +\left\vert 1,1\right\rangle _{\beta }\left\langle
1,1\right\vert )/2$. Calculations show that the maximal probability for Bob
to learn $q_{i}$ correctly is $\cos ^{2}(\pi /8)\simeq 0.8536$. Also, from
Fig.1 we know that type $b$ lies and honest results stand a higher chance to
be included in set $U$ than in set $N$, while type $c$ lies do the opposite.
However, step (10) shows that $c$ depends on the parity of the number of $1$%
's in $q_{i}$\ ($i\in N$). Bob's wrong guess on a single $q_{i}$\ value
could completely change this parity. Therefore, Bob cannot rely on the
probabilistic guess on $q$ and $N$. If he cannot learn the complete string $%
q $ precisely, the relationship between his guess on the parity of $%
\sum\nolimits_{i\in N}q_{i}$ and the actual value will be completely random,
so that he cannot determine the correct $f$ value to announce in step (7)
that leads to his desired $c$.

The only exception is type $b$ lies, as they never present in set $N$.
Therefore once Bob tells a type $b$ lie and Alice does not announce it as an
element of set $L$ in step (5), Bob knows with certainty that the
corresponding $i$ has to be included in set $U$, and he knows this $q_{i}$
from his own $q_{i}^{\prime \prime }$ as they are equal. So we must limit
the total number of type $b$ lies in the protocol. In step (4) Alice checks $%
\left\vert M\right\vert >d+s/4$\ which indicates $2d/s<f_{a}+f_{c}$. As Eqs.
(\ref{sizeN}) and (\ref{sizeU}) show that the total number of type $b$ lies
in $N\cup U$ is $s_{b}\equiv 3f_{b}/4$, we have%
\begin{eqnarray}
s_{b} &=&(\left\vert N\right\vert +\left\vert U\right\vert )-(f_{h}+\frac{1}{%
2}f_{a}+\frac{3}{4}f_{c})s  \nonumber \\
&\leq &(\left\vert N\right\vert +\left\vert U\right\vert )-\frac{1}{2}%
(f_{a}+f_{c})s  \nonumber \\
&<&(\left\vert N\right\vert +\left\vert U\right\vert )-d.
\end{eqnarray}%
Also, since set $L$ is announced in step (5), Bob knows every $q_{i}$ ($i\in
L$). Therefore, the total number of $q_{i}$'s that Bob knows exactly is $%
s_{b}+\left\vert L\right\vert <s-d$. This result means that there are more
than $d$ bits of the $s$-bit string $q$ remaining uncertain to Bob. From the
definition of set $C$ in step (1) we can see that knowing less than $s-d$
bits of $q$ is insufficient for Bob to determine which codeword it is. That
is, the requirement $\left\vert M\right\vert >d+s/4$\ guarantees that Bob
does not have enough information on $q$ in step (7) to determine the parity
of $\sum\nolimits_{i\in N}q_{i}$. The operations after step (7) are merely
the security checks against Alice's cheating. There is no more operation for
Bob to affect the coin-flip outcome. Thus we see that the protocol is
unconditionally secure against dishonest Bob.

Now consider Alice's cheating. Note that Bob's lying frequencies $f_{a}$, $%
f_{b}$ and $f_{c}$\ are never announced to Alice directly throughout the
protocol. If Alice follows algorithm III honestly, the sizes of her
announced sets $U$, $L$, $N$ will meet\ Eqs. (\ref{sizeU}), (\ref{sizeL})
and (\ref{sizeN}) automatically. But if she does not prepare the initial
states in the form of Eq. (\ref{eq2}), while she may still find a set $L$
with the proper size, she has trivial probability to make a close guess on
Bob's $f_{a}$, $f_{b}$, $f_{c}$\ (even though a small amount of statistical
fluctuation is allowed) and announced sets $U$, $N$ with the correct size.
Then she will be caught in step (9.2).

On the other hand, suppose that she prepared the states following Eq. (\ref%
{eq2}), but after Bob announced $\left\vert p_{i}^{\prime \prime
},q_{i}^{\prime \prime }\right\rangle _{\beta }$ in step (3), she does not
like the relationship between $q$ and $q^{\prime \prime }$ as they will not
lead to her desired $c$ value. Then to alter $c$, she needs to announce some
elements of $N$ as the elements of $U$ instead, or vice versa. But it is
well-known that entanglement cannot be created locally. Once Alice measured $%
\alpha _{i}$ in $\alpha _{i}\otimes \beta _{i}$ (i.e., $i\in M$), it is no
longer entangled with $\beta _{i}$. If she tries to claim that it is an
element of the unmeasured set $U$, there is a non-trivial probability that
in step (9.4) when Bob measures $\alpha _{i}$ and compares it with his
measurement on $\beta _{i}$, he will find a non-correlated result. For
example, suppose that for a certain $i$, Bob has announced $q_{i}^{\prime
\prime }=0$, and in step (8) Alice claims that $i\in U$ which implies that $%
q_{i}=q_{i}^{\prime \prime }=0$\ and $\alpha _{i}$ has not been measured. If
Bob has measured the corresponding $\beta _{i}$ and the actual result is $%
\left\vert p_{i}^{\prime },q_{i}^{\prime }\right\rangle _{\beta }=\left\vert
0,0\right\rangle _{\beta }$, then from Eq. (\ref{basis0}) he is expecting
that $\alpha _{i}$ has collapsed to the state $(\cos \theta _{i}\left\vert
x\right\rangle _{\alpha }+(\sin \theta _{i}/\sqrt{2})\left\vert
y\right\rangle _{\alpha })/n_{\alpha i}$. Here $n_{\alpha i}$\ is the
normalization constant, and we fixed $\theta _{i}=\pi /4$\ in the protocol.
To check Alice's announcement, in step (9.4) Bob can measure this $\alpha
_{i}$ in the basis $\{(\cos \theta _{i}\left\vert x\right\rangle _{\alpha
}+(\sin \theta _{i}/\sqrt{2})\left\vert y\right\rangle _{\alpha })/n_{\alpha
i},\overline{(\cos \theta _{i}\left\vert x\right\rangle _{\alpha }+(\sin
\theta _{i}/\sqrt{2})\left\vert y\right\rangle _{\alpha })/n_{\alpha i}}\}$,
where $\overline{(\cos \theta _{i}\left\vert x\right\rangle _{\alpha }+(\sin
\theta _{i}/\sqrt{2})\left\vert y\right\rangle _{\alpha })/n_{\alpha i}}$
denotes the state orthogonal to $(\cos \theta _{i}\left\vert x\right\rangle
_{\alpha }+(\sin \theta _{i}/\sqrt{2})\left\vert y\right\rangle _{\alpha
})/n_{\alpha i}$. If dishonest Alice already measured $\alpha _{i}$ in the
basis $\{\left\vert x\right\rangle _{\alpha },\left\vert y\right\rangle
_{\alpha }\}$ or $\left\vert \alpha _{i}\otimes \beta _{i}\right\rangle $\
was not prepared as Eq. (\ref{eq2'}), then Bob has a non-trivial probability
to find the measurement result as $\overline{(\cos \theta _{i}\left\vert
x\right\rangle _{\alpha }+(\sin \theta _{i}/\sqrt{2})\left\vert
y\right\rangle _{\alpha })/n_{\alpha i}}$ and thus detect her cheating.
Specially, if Bob has left a portion of $\beta _{i}$'s unmeasured before
step (9.4) (note that we elaborated before algorithm I that this will not
prevent Bob from controlling the values of $f_{a}$, $f_{b}$ and $f_{c}$) and
the current $\beta _{i}$ happens to belong to this portion, then he can
perform a collective measurement on $\alpha _{i}\otimes \beta _{i}$\ to see
whether they can be projected to the state in Eq. (\ref{eq2'}).

Alice's announcing an element of $U$ as that of $N$ is also detectable. For $%
\forall i\in U$, Bob's announced $\left\vert p_{i}^{\prime \prime
},q_{i}^{\prime \prime }\right\rangle _{\beta }$ satisfies $q_{i}^{\prime
\prime }=q_{i}$. Take $q_{i}=0$\ for example. If Bob announced $\left\vert
p_{i}^{\prime \prime },q_{i}^{\prime \prime }\right\rangle _{\beta
}=\left\vert 0,0\right\rangle _{\beta }$, Eq. (\ref{basis0}) shows that $%
\alpha _{i}$ should collapse to $(\cos \theta _{i}\left\vert x\right\rangle
_{\alpha }+(\sin \theta _{i}/\sqrt{2})\left\vert y\right\rangle _{\alpha
})/n_{\alpha i}$ if $\left\vert p_{i}^{\prime \prime },q_{i}^{\prime \prime
}\right\rangle _{\beta }$ is not a lie. So if Alice measures $\alpha _{i}$
in the basis $\{(\cos \theta _{i}\left\vert x\right\rangle _{\alpha }+(\sin
\theta _{i}/\sqrt{2})\left\vert y\right\rangle _{\alpha })/n_{\alpha i},%
\overline{(\cos \theta _{i}\left\vert x\right\rangle _{\alpha }+(\sin \theta
_{i}/\sqrt{2})\left\vert y\right\rangle _{\alpha })/n_{\alpha i}}\}$ and the
result is $\overline{(\cos \theta _{i}\left\vert x\right\rangle _{\alpha
}+(\sin \theta _{i}/\sqrt{2})\left\vert y\right\rangle _{\alpha })/n_{\alpha
i}}$, she knows that Bob lies. But if Bob tells a type $b$ lie, i.e., his
actual result is $\left\vert p_{i}^{\prime },q_{i}^{\prime }\right\rangle
_{\beta }=\left\vert \urcorner p_{i}^{\prime \prime },q_{i}^{\prime \prime
}\right\rangle _{\beta }=\left\vert 1,0\right\rangle _{\beta }$, Eq. (\ref%
{basis1}) shows that $\alpha _{i}$ actually collapsed to $((\cos \theta _{i}/%
\sqrt{2})\left\vert x\right\rangle _{\alpha }+\sin \theta _{i}\left\vert
y\right\rangle _{\alpha })/n_{\alpha i}^{\prime }$. It has a non-trivial
probability to be detected as $(\cos \theta _{i}\left\vert x\right\rangle
_{\alpha }+(\sin \theta _{i}/\sqrt{2})\left\vert y\right\rangle _{\alpha
})/n_{\alpha i}$ since they are not orthogonal. That is, type $b$ lies
cannot be detected with probability $100\%$ when $q_{i}^{\prime \prime
}=q_{i}$, in contrast to the case $q_{i}^{\prime \prime }\neq q_{i}$.
Consequently, if Alice picks an element of $U$ and wants to claim that it
belongs to $N$ instead, Eq. (\ref{sizeU}) shows that she stands probability $%
3f_{b}/4$\ to come across a type $b$ lie, and she cannot always distinguish
it even if she measures $\alpha _{i}$. As it was shown in algorithm III that
$N$ should not contain any type $b$ lies, Alice's claiming $i\in N$ will
immediately be caught as cheating.

Thus it is shown that either Alice announces a single element of $N$ as that
of $U$, or vice versa, she stands a non-trivial probability $\varepsilon $
to be detected. More importantly, while altering one single element of $N$
and $U$ (i.e., changing one single $q_{i}$\ value) could be sufficient for
changing the parity of $\sum\nolimits_{i\in N}q_{i}$ and thus affect the
final coin-flip outcome $c$, in our protocol the string $q$ is required to
be a codeword of $C$. As the minimal distance between codewords is $d$,
changing one or few $q_{i}$'s of a codeword will result in $q\notin C$ and
be detected in step (9.1). Consequently, dishonest Alice has to alter at
least $d/2$ bits of $q$ (suppose that she started with a $q$ lying half way
between two codewords) to cheat successfully. But then the total probability
for escaping the detection will be at the order of magnitude of $%
(1-\varepsilon )^{d/2}$. By choosing a very high $s$ value in our protocol,
the $d$ value can also be increased, so that Alice's probability for
successfully biasing the coin-flip result can be made arbitrarily close to
zero. Thus our protocol is also unconditionally secure against dishonest
Alice.

\section{Relationship with the no-go proofs}

Currently there are many references on the no-go proofs of unconditionally
secure QCF \cite{qbc121}, \cite%
{qi58,qbc46,qi145,qi146,qi76,qi817,qbc37,qbc54,qi246,qbc132,qbc19,qbc106,qbc112}%
. Among them, Ref. \cite{qbc121} is a review on the previous results of
Refs. \cite{qi817,qbc37,qbc19} (which we will discuss later in this section), without supplying any new proof of its own.
Ref. \cite{qbc46} provided the specific cheating strategy on the protocol in
Ref. \cite{qi132} only. Ref. \cite{qi146} is merely the specific cheating
strategy on a class of cheat-sensitive protocols proposed in Ref. \cite%
{qi148}. The no-go proofs in Refs. \cite{qi76,qbc112} are aimed at the QCF
protocols built upon QBC. But we already elaborated in Refs. \cite%
{HeJPA,HeQIP,HeBook,HePRSA} that the no-go proofs on unconditionally secure
QBC is not sufficiently general to cover all protocols.
Refs. \cite{qbc54,qi246} studied a family of weak QCF protocols which are
based on the $n$-coin-games defined in Ref. \cite{qi246}. Ref. \cite{qbc106}
studied the attack on two specific practical QCF protocols. That is, they
are not supposed to be general.

The rest no-go proofs \cite{qi58,qi145,qi817,qbc37,qbc132,qbc19} were
claimed to apply to all kinds of QCF protocols. Thus our current result
seems to conflict with them, and it is natural to question which points in
their reasoning do not fit our protocol. As each of these proofs has its own
approach, it is hard to answer to them all in one common sentence.
Therefore, we will explain below how our protocol evades these no-go proofs
one by one.

Ref. \cite{qi58}: This reference proved that ideal QCF (i.e., the bias $%
\varepsilon $ equals to $0$ precisely, instead of being arbitrarily close to
$0$) is impossible. As seen from the proof of its Theorem 2 in page 8, their
approach is to argue that if an ideal QCF can be done in $N$ rounds of
communication, then it can be done in $N-1$ rounds. By repeating this
induction, it seems that QCF can be done without any communication -- which
is an absurd result that can easily be disproved. But this no-go proof does
not apply to our protocol for the following two reasons. First, ours is a
non-ideal QCF in the sense that if either Alice or Bob wants to bias the
final coin flipping result, the successful probability can be made
arbitrarily small with the increase of the parameter $s$\ (which allows
higher $k$ and $d$ values while keeping $k/s$\ and $d/s$ as constants), but it
does not vanish rigorously as long as $s$ remains finite. This is equivalent
to having an arbitrarily small but non-vanishing bias $\varepsilon $, which
does not fit the definition of ideal QCF. As stated clearly in the last
paragraph of section 3 of Ref. \cite{qi58}, their proof cannot be
generalized to such a non-ideal case. Secondly and more importantly, the $N$
to $N-1$ rounds induction obviously cannot apply to our protocol. For
example, step (9) of our protocol is a security check. If it is removed,
then the protocol surely becomes an insecure QCF, which is not equivalent to
the original one. Also, in step (8) Alice announces sets $N$ and $U$\ which
is necessary for Bob to calculate the coin-flip result $c$. If it is also
removed, then Bob cannot get $c$ so that the protocol is not a complete QCF. Thus, we
can see that the induction approach not only fails to work in our case, but
also fails to cover basically any serious protocol, because removing each
round of communication from a protocol one by one will eventually remove the
security check. A protocol without security checks is surely insecure. As a
result, the induction used in this no-go proof is actually a pointless
approach since it only works for naive protocols which contain no security
check at all.

Ref. \cite{qi145}: This paper showed that to achieve a bias of at most $%
\varepsilon $, a QCF protocol must use at least $\Omega (\log \log
(1/\varepsilon ))$ rounds of communication. According to the proof of its
Lemma 12 on page 14, \textquotedblleft to bias the coin towards $0$, Alice
just runs the honest protocol with her starting state being $\left\vert \psi
_{A}^{\prime }\right\rangle $\ instead of $\left\vert \psi _{A}\right\rangle
$\textquotedblright . Here $\left\vert \psi _{A}\right\rangle $ denotes the
initial state at Alice's side when the protocol is executed honestly, $%
\left\vert \psi _{A}^{\prime }\right\rangle $\ is the remaining state when
Alice applies $M$ on $\left\vert \psi _{A}\right\rangle $ and gets the
coin-flip outcome $0$, with $M$ being the best measurement that
distinguishes the two density matrices corresponding to the two outcomes $0$
and $1$, respectively. But in our protocol, such a measurement $M$ (not to
be confused with set $M$) will be determined by Bob's choice of $q^{\prime
\prime }$ in step (3). Alice does not know it at the beginning of the
protocol. Therefore, at this stage she cannot know what would be the form of
$\left\vert \psi _{A}^{\prime }\right\rangle $, so that this cheating
strategy cannot be implemented. That is, the no-go proof in Ref. \cite{qi145}
actually bases on the assumption that the measurement $M$ is always known to
Alice, though the author did not aware nor state explicitly that it is an
assumption. Now, by mixing Bob's secret information into the construction of
the measurement $M$, our protocol managed to present a counterexample
showing that this assumption does not always hold in every QCF protocol.

Ref. \cite{qi817}: This is known as the very first proof which lowered the
bound to $\varepsilon \geq 1/\sqrt{2}-1/2$. While this reference was widely
cited, what can be found online is merely a scan of the slides. Thus the
details of the proof is inaccessible. Fortunately, its result was reproduced
in Ref. \cite{qbc37}. Therefore, we will analyze it based on Ref. \cite{qbc37}
below.

Ref. \cite{qbc37}: According to its Definition 8, any QCF protocol is
treated as a series of Alice's and Bob's unitary transformations $U_{A,j}$'s
and $U_{B,j}$'s ($1\leq j\leq N$) on the Hilbert space, and their final
measurements $\Pi _{A,c}$ and $\Pi _{B,c}$ for finding the coin-flip outcome
$c=0,1$ while the protocol does not abort. Its Lemmas 10 and 11 suggested
that the optimal strategy of Bob trying to force outcome $1$ is the solution
to the primal semidefinite program (SDP) in its Eqs. (29)-(31), whose dual
SDP is given by its Eqs. (32)-(34). Note that its Eqs. (29), (31), (33) and
(34) all depend on Alice's $U_{A,j}$'s and $\Pi _{A,1}$. In our protocol,
Alice's choices of the $q$ value in step (2) play the role of $U_{A,j}$'s,
which is unknown to Bob before he needs to announces $q^{\prime \prime }$ in
step (3) and $f$ in step (7). Since the coin-flip result $c$ is determined
by $f$ and the comparison between $q$ and $q^{\prime \prime }$, once Bob
announced $q^{\prime \prime }$ and $f$, $c$ is fixed so that he can no
longer alter it. Therefore, though the optimal SDP that leads to the
cheating probability $p_{\ast 1}\geq 1/\sqrt{2}$ (i.e., the Kitaev's bound $%
\varepsilon \geq 1/\sqrt{2}-1/2$) exists, Bob does not have enough
information on $U_{A,j}$'s to compute it before the coin-flip result is
generated. The same analysis also applies to dishonest Alice. In short, the
problem of this no-go proof is that it studies QCF with a static point of
view: Alice and Bob always know all data of the other party. It fails to
notice that in some protocols (including ours), by arranging the steps
wisely, it is definitely possible to control what information is known to
each party at different stages of the protocol, so that proving the
existence of the SDP does not imply that the protocol is proven insecure.
The cheater may not know the SDP by the time he needs to apply it, and when
he finally gets to know it, cheating will be all too late. This result is
very similar to the case of Ref. \cite{qbc132} below, where the optimal
cheating strategy exists but the cheater cannot reach it.

Ref. \cite{qbc132}: This proof recovered Kitaev's bound $\varepsilon \geq 1/%
\sqrt{2}-1/2$ for strong QCF \cite{qi817} with a different presentation. In
brief, as shown in its page 8, suppose that $\{A_{0},A_{1},A_{abort}\}$ is
honest-Alice's strategy and $\{B_{0},B_{1},B_{abort}\}$ is honest-Bob's
co-strategy in a QCF protocol, corresponding to the outcomes $0$, $1$ and $%
abort$, respectively. The definition of QCF implies%
\begin{equation}
\frac{1}{2}=\left\langle A_{0},B_{0}\right\rangle =\left\langle
A_{1},B_{1}\right\rangle .
\end{equation}%
Let $p$ be the maximum probability that a cheating Bob can force
honest-Alice to output a fixed $c\in \{0,1\}$. Then its Theorem 9 implies
that there must exist a strategy $Q$ for Alice such that $A_{c}\leq pQ$. If
a cheating Alice plays this strategy $Q$, then honest-Bob outputs $c$ with
probability%
\begin{equation}
\left\langle Q,B_{c}\right\rangle \geq \frac{1}{p}\left\langle
A_{c},B_{c}\right\rangle =\frac{1}{2p}.  \label{eq132}
\end{equation}%
Given that%
\begin{equation}
\max \left\{ p,\frac{1}{2p}\right\} \geq \frac{1}{\sqrt{2}}
\end{equation}%
for all $p>0$, either honest-Alice or honest-Bob can be convinced to output $%
c$ with probability at least $1/\sqrt{2}$, so that the bias satisfies $%
\varepsilon \geq 1/\sqrt{2}-1/2$.

Now consider our protocol. First, let us suppose that Alice is honest. Then
the quantum states that Bob received are his halves of the entangled states
faithfully prepared as our Eq. (\ref{eq2'}). As a result, Bob cannot learn the
exact values of Alice's $q_{i}$'s\ from his own measurement, since any value
of $\left\vert p_{i}^{\prime },q_{i}^{\prime }\right\rangle _{\beta }$\ is
possible no matter $q_{i}=0$\ or $q_{i}=1$. %
%
%
Also, $c$ is determined by the parity of $\sum\nolimits_{i\in N}q_{i}$,
which depends sensitively on the value of every single $q_{i}$.
Consequently, making a probabilistic guess of $q_{i}$ is useless for Bob to
alter $c$. Therefore, according to the above description of Ref. \cite%
{qbc132}, our protocol has $p=1/2$. Then Eq. (\ref{eq132}) indicates that
there is a cheating strategy $Q$ which can maximize $\left\langle
Q,B_{c}\right\rangle $ to $1$, which is Alice's probability for forcing the
output.

However, the question is whether Alice knows this $Q$. Note that the task of
$Q$ is not merely to produce Alice's desired $c$ value, but also to bring her
through the security checks successfully. In our protocol, to obtain the
output $c$, Bob's strategies are not limited to one single $B_{c}$. Even if
he always announced the same values of $\left\vert p_{i}^{\prime \prime
},q_{i}^{\prime \prime }\right\rangle _{\beta }$'s and $f$ in steps (3) and
(7), his choice on the actual measurement bases $p^{\prime }\equiv
p_{1}^{\prime }p_{2}^{\prime }...p_{s}^{\prime }$ can be different,
resulting in different locations and types of his lies. Also, in step (9.4)
Bob has the freedom on choosing the measurement bases $p^{\prime (\alpha
)}\equiv \bigotimes\nolimits_{i\in U}p_{i}^{\prime (\alpha )}$ for checking $%
\alpha _{i}$'s in set $U$. That is, even for the same $\left\vert
p_{i}^{\prime \prime },q_{i}^{\prime \prime }\right\rangle _{\beta }$'s and $%
f$, he still has many different strategies, each of which is corresponding to
a different choice of $p^{\prime }$ and $p^{\prime (\alpha )}$, so we can
denote it as $B_{c}(p^{\prime },p^{\prime (\alpha )})$. While $p^{\prime }$
and $p^{\prime (\alpha )}$ do not affect $c$, they determine what kinds of
Alice's states can pass Bob's security checks. Therefore, Alice strategy $Q$
has to match Bob's $B_{c}(p^{\prime },p^{\prime (\alpha )})$, so we can
denote it as $Q(B_{c}(p^{\prime },p^{\prime (\alpha )}))$.

As each $p_{i}^{\prime }$ has two possible values, there are totally $2^{s}$%
\ different choices for $p^{\prime }$, leading to at least $2^{s}$\
different strategies $B_{c}(p^{\prime },p^{\prime (\alpha )})$. The choices
for $p^{\prime (\alpha )}$ make the number of $B_{c}(p^{\prime },p^{\prime
(\alpha )})$ even higher. Now, recall that in our protocol $p^{\prime }$\
and $p^{\prime (\alpha )}$\ are never required to be announced to Alice.
Consequently, although the optimal cheating strategy $Q(B_{c}(p^{\prime
},p^{\prime (\alpha )}))$ that maximize $\left\langle Q(B_{c}(p^{\prime
},p^{\prime (\alpha )})),B_{c}(p^{\prime },p^{\prime (\alpha
)})\right\rangle $ could exist, Alice has more than $2^{s}$\ different $%
Q(B_{c}(p^{\prime },p^{\prime (\alpha )}))$\ to choose from, but she does not
know which is the optimal one, since she does not know $p^{\prime }$\ and $%
p^{\prime (\alpha )}$. Thus, she only has less than probability $1/2^{s}$ to
find the optimal $Q(B_{c}(p^{\prime },p^{\prime (\alpha )}))$\ and reach the
maximal bias described in Eq. (\ref{eq132}). That is why the proof in Ref.
\cite{qbc132} does not apply to our protocol.

Two lessons can be learned from the failure of this no-go proof. (i) Like
the proof in Ref. \cite{qbc37}, it did not notice the difference between
existence and accessibility. It is insufficient to deny the security of a
protocol by merely proving the theoretical existence of an optimal strategy
without proving that it is accessible to the cheater. (ii) Recall the famous
quote from Gilbert K. Chesterton: \textquotedblleft Where does a wise man
hide a leaf? In the forest. But what does he do if there is no forest? He
grows a forest to hide it in.\textquotedblright\ This is exactly what
happens in our protocol. By introducing more data, more freedom of choices
and more uncertainty, the relationship between the possible strategies of
Alice and Bob is no longer as simple as considered in Ref. \cite{qbc132}.
Bob can flood Alice with an excessive amount of choices so that the optimal
one gets buried before her eyes.

%
%
%
%

Ref. \cite{qbc19}: This reference studied the bounds for both strong and
weak QCFs. The no-go result of strong QCF is presented in its Theorem 2,
which was proven based on its Lemmas 10 and 11. As clearly stated in its
section 4.2, Lemma 11 was obtained directly from Ref. \cite{qbc37}. Since we
already elaborated above why the proof in Ref. \cite{qbc37} does not apply
to our protocol, it immediately follows that the result of Ref. \cite{qbc19}
also fails for exactly the same reason. Moreover, there is another problem
worth pointing out which may hurt the generality of this no-go proof, even
though it does not relate directly with our protocol. That is, footnote 4
on its page 3 is incorrect. It said that \textquotedblleft the players can
always add a final round to check if they have the same value and output the
dummy symbol if the values differ\textquotedblright . But if there is a
secure QCF protocol without this round, then adding such a round will make
it insecure, because a dishonest party will be able to bias the output by
always announcing the value he preferred as what he obtained from the
protocol, regardless what is the actual output value. Whenever his preferred
value differs from the actual one, the protocol will abort so he has nothing
to lose. Therefore, dummy output in QCF should not be allowed to be
generated in the way described in this footnote.

All in all, we pointed out in the above paragraphs that each of the no-go
proofs in Refs. \cite{qi58,qi145,qi817,qbc37,qbc132,qbc19} has its
particular problems, so that it actually proved the insecurity of a certain class of
QCF protocols only, despite that they were claimed to be general. Here we
would like to further address a common issue that widely exists not only in
the no-go proofs of QCF, but also in those of QBC and QOT. That is, the use
of Yao's model \cite{qi75} (as called by Ref. \cite{qi58}). In this model,
any two-party quantum protocol is reduced to the following form: Alice
and Bob share a quantum system in the Hilbert space $H_{A}\otimes H_{B}$\
(where $H_{A}$\ ($H_{B}$) is Alice's (Bob's) Hilbert space); they interact
with each other through quantum channels only, without any classical
communication; they apply unitary (therefore reversible) transformations on
the quantum system in turns, and perform measurements only at the end of the
protocol. Note that Yao did not claim the model to be general, but Ref. \cite%
{qi58} claimed that \textquotedblleft any realistic two-party computation
can be described by Yao's model\textquotedblright . There are two important
points in the model:

\textit{(i) All classical communications can be replaced by quantum ones.}

\textit{(ii) Measurements can always be delayed until the very end.}

As can be seen in Ref. \cite{qi105}, their reasoning behind point (i) is
that classical communications are a special case of quantum communications.
For example, when Alice wants to send a classical bit $k$ ($k\in \{0,1\}$)
to Bob, rather than using classical communications, she can send Bob a
quantum state $\left\vert k\right\rangle $\ instead. If Bob measures the
state received from Alice in the computational basis $\{\left\vert
0\right\rangle ,\left\vert 1\right\rangle \}$, he can recover the classical
information of Alice. The reasoning behind point (ii) is as follows. Suppose
that Alice is required to send Bob either the pure state $\left\vert
0\right\rangle _{\beta }$\ or $\left\vert 1\right\rangle _{\beta }$ with the
equal probabilities. Now if Alice prepares the entangled state $\left\vert
\alpha \otimes \beta \right\rangle =(\left\vert x\right\rangle _{\alpha
}\otimes \left\vert 0\right\rangle _{\beta }+\left\vert y\right\rangle
_{\alpha }\otimes \left\vert 1\right\rangle _{\beta })/\sqrt{2}$\ and sends
Bob the second qubit $\beta $ (whose state is in a mixture), Bob cannot tell
the difference via his own measurement. Consequently, Alice can keep her own
qubit $\alpha $ unmeasured (so that her choice on the state of $\beta $ remains undetermined), until she needs to unveil all secret data at the
end of the protocol.

These two points were challenged (e.g., see Ref. \cite{qi360}) during the
early years of the no-go proofs, but became widely accepted later on, since
most challengers cannot come up with unconditionally secure protocols as
convincing counterexamples. But here we will show that they are indeed
illogical. Consider point (i) first. While there is no doubt that classical
communications are a special case of quantum communications, it does not
mean that they are equivalent and exchangeable. According to the quantum
no-cloning theorem, arbitrary unknown quantum
states cannot be cloned, while there is no such a no-cloning theorem for
classical informations. Replacing classical communications with quantum ones
actually ignores the specialty of the former. In the above reasoning behind
point (i), Alice sends Bob the quantum state $\left\vert k\right\rangle $\
when she is required to send the classical bit $k$. It is true that \textit{if} Bob
measures $\left\vert k\right\rangle $\ in the basis $\{\left\vert
0\right\rangle ,\left\vert 1\right\rangle \}$, there will be no difference
than sending $k$ via classical communications. But what if Bob measures $%
\left\vert k\right\rangle $\ in another basis such as $\{\left\vert \pm
\right\rangle \equiv (\left\vert 0\right\rangle \pm \left\vert
1\right\rangle )/\sqrt{2}\}$? More generally, if one party is required to
send many classical bits but he replaces them all with quantum states, the
other party can perform collective measurements on them together instead of
measuring each of them in the computational basis. This could enable him to
achieve some global information (e.g., parity or weight of the bit-strings) while
causing less disturbance on the quantum states than individual measurements.
If Alice sent the bits using classical communications, there will be no such
alternative measurements. But replacing them with quantum communications
will open up these alternatives, which makes room for potential cheatings.
Thus, we can see that using measurements can reveal the inequivalence between
classical communications and quantum ones.

But now the supporters of the no-go proofs may reach out for point (ii),
saying that you cannot force measurements until the end, so that the
inequivalence between classical and quantum communications can remain
unexposed. Now let us show that point (ii) is also faulty. Consider that
Alice prepares the entangled state $\left\vert \alpha \otimes \beta
\right\rangle =(\left\vert x\right\rangle _{\alpha }\otimes \left\vert
0\right\rangle _{\beta }+\left\vert y\right\rangle _{\alpha }\otimes
\left\vert 1\right\rangle _{\beta })/\sqrt{2}$\ and sends Bob the second
qubit $\beta $. Indeed, Bob cannot tell it apart from the pure state $%
\left\vert 0\right\rangle _{\beta }$\ or $\left\vert 1\right\rangle _{\beta
} $ via his own measurement alone. But what if Bob says classically that
\textquotedblleft I measured $\beta $ and the result is $\left\vert
0\right\rangle _{\beta }$\textquotedblright\ and asks Alice to judge whether
he is lying? Surely Alice can no longer keep her qubit $\alpha $ unmeasured,
otherwise she has absolutely no idea whether Bob lied or not. Thus, we show
that measurements can be forced to complete without delay by using classical
communications.

However, the supporters of the no-go proofs could turn back to point (i) in
this case, saying that it is known that there could be no classical
communications in the protocol. Then the argument would restart all over
again. Therefore, we can see that points (i) and (ii) actually form an
interesting loop. If you agree the validity of point (i), then point (ii)
can be proven valid. Taking point (ii) as true, then point (i) can be proven
too. While the situation might appears self-consistent, in fact it is
completely the opposite. That is, if we do not accept point (i), then point
(ii) fails. Without point (ii), point (i) will no longer hold either. By
carefully designing what classical communications to use in the protocol, it
is possible to break both points simultaneously. The Yao's model used to
work for some cases, because in the corresponding protocols when one party
sends $\left\vert k\right\rangle $\ to represent a classical bit $k$, the
other party always measures it in the basis $\{\left\vert 0\right\rangle
,\left\vert 1\right\rangle \}$ honestly without considering other
alternative bases, so that point (i) is made valid. For instance, in the
well-known quantum key distribution (QKD) protocols, Alice and Bob always
collaborate honestly in order to defeat the attack from external
eavesdroppers, which is not the legitimate participants of the protocol. Both
Alice and Bob have no motivation to cheat, so that there is no problem to
study QKD protocols using Yao's model. But in two-party computation
cryptographies such as QCF and QBC, there is a significant difference. The
possible cheatings come from the internal legitimate participants Alice and Bob. They
do not fully trust each other, so that we can no longer assume that they
always collaborate. In conclusion, while Yao's model may work well for
cryptographic tasks like QKD where all legitimate participants always
collaborate honestly against external attacks, it is not sufficiently
general to cover cryptographic tasks where the internal legitimate participants
could cheat, e.g., QCF, QBC as well as QOT. Despite that most no-go proofs
of QCF did not mention Yao's model explicitly, they actually used the same
approach, so that they all have the same limitation.

Finally, it is worth noting that Kitaev proposed a reverse version of point
(i). He suggested that all quantum communications in a protocol can be
replaced by classical ones \cite{qi817}. The reason is that Alice and Bob
can share a sufficiently large number of entangled states before the
protocol starts. When they need to transmit quantum states later in the
protocol, they use quantum teleportation \cite{qi179} instead, which
involved local operations and classical communications (LOCC) at this stage
only. However, we should notice that in quantum teleportation, the sender
needs to announce the result of his local measurement honestly to the
receiver, so that the latter can perform the correct unitary transformation
to recover the transmitted quantum information from his share of the
entangled state. In cryptographic tasks like QCF where the potential
cheatings come from the internal legitimate participants, there is no guarantee
that the sender will always announce this information honestly. Therefore,
Kitaev's approach cannot be considered sufficiently general either.

\section{Discussions}

In summary, we proposed a QCF protocol, and showed that it has features
which are not covered in all no-go proofs. Thus it can break the constraint
of existing security bounds on QCF.

Note that our proposed protocol and all the above security analysis were
presented with an ideal setting in mind, where practical imperfections of
experimental devices are not considered, e.g., transmission errors,
detection loss or dark counts. This is because the current work is focused
on the fundamental theoretical problem whether quantum mechanics allows
unconditionally secure QCF. If taking these practical imperfections into
account, the security analysis will be too complicated to check, and may
even be led to a wrong direction. We may consider how to adjust our protocol
to make it comply with the imperfect noisy environment in future works
though, after the debate on the theoretical existence of unconditionally
secure QCF is settled.

\bigskip

\noindent \textbf{Funding} This work was supported in part by Guangdong
Basic and Applied Basic Research Foundation (Grant No. 2019A1515011048).

\bigskip

\noindent \textbf{Conflict of interest} The author declares no conflict of
interest.


\bigskip

\begin{thebibliography}{99}
\bibitem{qi365} Bennett C H, Brassard G. Quantum cryptography: public key
distribution and coin tossing., In: Proceedings of IEEE International
Conference on Computers, Systems, and Signal Processing. New York: IEEE,
1984. 175

\bibitem{qi150} Hardy L, Kent A. Cheat sensitive quantum bit commitment.
Phys Rev Lett, 2004, 92: 157901

\bibitem{qi442} Ishizaka S. Dilemma that cannot be resolved by biased
quantum coin flipping. Phys Rev Lett, 2008, 100: 070501

\bibitem{HeSR} He G P. Security bound of cheat sensitive quantum bit
commitment. Sci Rep, 2015, 5: 9398

\bibitem{qbc121} Broadbent A, Schaffner C. Quantum cryptography beyond
quantum key distribution. Design Code Cryptogr, 2016, 78: 351

\bibitem{qi139} Kilian J. Founding cryptography on oblivious transfer. In:
Proceedings of 1988 ACM Annual Symposium on Theory of Computing. New York:
ACM, 1988. 20

\bibitem{qi58} Lo H -K, Chau H F. Why quantum bit commitment and ideal
quantum coin tossing are impossible. Physica D, 1998, 120: 177

\bibitem{qbc46} Leslau B. Attacks on symmetric quantum coin-tossing
protocols. arXiv:quant-ph/0104075v2

\bibitem{qi145} Ambainis A. A new protocol and lower bounds for quantum coin
flipping. arXiv:quant-ph/0204022v1

\bibitem{qi146} Ambainis A. Lower bound for a class of weak quantum coin
flipping protocols. arXiv:quant-ph/0204063v1

\bibitem{qi76} Nayak A, Shor P. Bit-commitment-based quantum coin flipping.
Phys Rev A, 2003, 67: 012304

\bibitem{qi817} Kitaev A. A negative result about quantum coin flipping.
Lecture delivered at the 2003 Annual Quantum Information Processing (QIP)
Workshop, Mathematical Sciences Research Institute, Berkeley, CA, 2003.
Available from:
http://www.msri.org/publications/ln/msri/2002/qip/
kitaev/1/meta/aux/kitaev.pdf

\bibitem{qbc37} Ambainis A, Buhrman H, Dodis Y, R\"{o}hrig H. Multiparty
quantum coin flipping. arXiv:quant-ph/0304112v2

\bibitem{qbc54} Mochon C. Quantum weak coin-flipping with bias of 0.192.
arXiv:quant-ph/0403193v2 

\bibitem{qi246} Mochon C. Large family of quantum weak coin-flipping
protocols. Phys Rev A, 2005, 72: 022341 

\bibitem{qbc132} Gutoski G, Watrous J. Toward a general theory of quantum
games. In: Proceedings of the Thirty-Ninth Annual ACM Symposium on Theory of
Computing (STOC 2007). New York: ACM, 2007. 565

\bibitem{qbc19} H\"{a}nggi E, Wullschleger J. Tight bounds for classical and
quantum coin flipping. arXiv:1009.4741v2

\bibitem{qbc106} Sajeed S, Radchenko I, Kaiser S, Bourgoin J -P, Pappa A,
Monat L, Legr\'{e} M, Makarov V. Attacks exploiting deviation of mean qubit
number in quantum key distribution and coin tossing. Phys Rev A, 2015, 91:
032326 

\bibitem{qbc112} Nayak A, Sikora J, Tun\c{c}el L. Quantum and classical
coin-flipping protocols based on bit-commitment and their point games.
arXiv:1504.04217v1

\bibitem{qi132} Mayers D, Salvail L, Chiba-Kohno Y. Unconditionally secure
quantum coin tossing. arXiv:quant-ph/9904078v1

\bibitem{qi148} Spekkens R W, Rudolph T. Quantum protocol for
cheat-sensitive weak coin flipping. Phys Rev Lett, 2002, 89: 227901

\bibitem{qbc29} Silman J, Chailloux A, Aharon N, Kerenidis I, Pironio S,
Massar S. Fully distrustful quantum bit commitment and coin flipping. Phys
Rev Lett, 2011, 106: 220501

\bibitem{qbc76} Yang Q, Ma J -J, Guo F -Z, Wen Q -Y. Semi-loss-tolerant
strong quantum coin-flipping protocol using quantum non-demolition
measurement. Quantum Inf Process, 2014, 13: 1537

\bibitem{qbc81} Pappa A, et al. Experimental plug and play quantum coin
flipping. Nat Commun, 2014, 5: 3717

\bibitem{qbc117} Zhang S, Zhang Y X. Quantum coin flipping secure against
channel noises. Phys Rev A, 2015, 92: 022313

\bibitem{qbc131} Zhao L Y, Yin Z Q, Wang S, Chen W, Chen H, Guo G C, Han Z
F. Measurement-device-independent quantum coin tossing. Phys Rev A, 2015,
92: 062327


\bibitem{HeJPA} He G P. Quantum key distribution based on orthogonal states
allows secure quantum bit commitment. J Phys A: Math Theor, 2011, 44: 445305

\bibitem{HeQIP} He G P. Simplified quantum bit commitment using single
photon nonlocality. Quantum Inf Process, 2014, 13: 2195

\bibitem{HeBook} He G P. Chapter 4: Density matrices in quantum bit
commitment. In: Danielsen N V, ed. Understanding Density Matrices. New York:
Nova Science Publishers, 2019. 139--164

\bibitem{HePRSA} He G P. Unconditionally secure quantum bit commitment based
on the uncertainty principle. Proc R Soc A, 2019, 475(2222): 20180543

\bibitem{qi75} Yao A C C. Security of quantum protocols against coherent
measurements. In: Proceedings of the 26th Symposium on the Theory of
Computing. New York: ACM, 1995. 67

\bibitem{qi105} Chau H F, Lo H -K. Making an empty promise with a quantum
computer. Fortsch Phys, 1998, 46(4--5): 507

\bibitem{qi360} Dang M -D. Two-party models and the no-go theorems.
arXiv:quant-ph/0608165

\bibitem{qi179} Bennett C H, Brassard G, Cr\'{e}peau C, Jozsa R, Peres A,
Wootters W K. Teleporting an unknown quantum state via dual classical and
Einstein-Podolsky-Rosen channels. Phys Rev Lett, 1993, 70: 1895
\end{thebibliography}
\end{document}